# IoT Data Analytics Using Deep Learning

Xiaofeng Xie, Di Wu, Siping Liu, Renfa Li

**Abstract**: Deep learning is a popular machine learning approach which has achieved a lot of progress in all traditional machine learning areas. Internet of thing (IoT) and Smart City deployments are generating large amounts of time-series sensor data in need of analysis. Applying deep learning to these domains has been an important topic of research. The Long-Short Term Memory (LSTM) network has been proven to be well suited for dealing with and predicting important events with long intervals and delays in the time series. LTSM networks have the ability to maintain long-term memory. In an LTSM network, a stacked LSTM hidden layer also makes it possible to learn a high level temporal feature without the need of any fine tuning and preprocessing which would be required by other techniques. In this paper, we construct a long-short term memory (LSTM) recurrent neural network structure, use the normal time series training set to build the prediction model. And then we use the predicted error from the prediction model to construct a Gaussian naive Bayes model to detect whether the original sample is abnormal. This method is called LSTM-Gauss-NBayes for short. We use three real-world data sets, each of which involve long-term time-dependence or short-term time-dependence, even very weak time dependence. The experimental results show that LSTM-Gauss-NBayes is an effective and robust model.

## 1. Introduction

In many areas, such as the natural sciences, social sciences, and engineering, phenomena are best analyzed with time-series data. In the field of Internet of things[1][2], time series data can be generated by weather stations, RFID tags, IT infrastructure components[3], and some other sensors[4], as shown in Fig. 1(a). In each business process and every application of the Internet of things sensor, the time series data can be used for process optimization or knowledge discovery. When carefully analyzed such data can reveal operational trends, patterns, variability, changes, covariation, cycle abnormalities, anomaly and abnormal value rate. Traditional time series processing technique uses a statistical indicator such as cumulative sum (CUSUM) and exponentially weighted moving average (EWMA) in a time window [5] to detect potential changes in the distribution. The length of the time window usually needs to be predetermined and the result is highly dependent on the parameter. In addition, there are some sequence models, such as conditional random field model, d Kalman filter, Markov model, dealing with sequential data but are ill-equipped to learn long-range dependencies. What's more, other models require domain knowledge or feature engineering, thus they provide fewer opportunities for accidental discovery. In contrast, neural network learning techniques allow unforeseen structures to be found. Recurrent neural network (RNN) [6] can theoretically solve long delayed tasks without requiring predefined time steps. However, because of the simplicity of the hidden layer units' structure, gradient explosion or vanishing gradient [7] is easy to occur over longer time series tasks. The long short-term memory (LSTM) neural network [8][9] is a

Xiaofeng Xie, Di Wu, Siping Liu and Renfa Li are with the Key Laboratory for Embedded and Networking Computing of Hunan Province, Hunan University.
Di Wu is the corresponding author (Email: dwu@hnu.edu.cn).

variant of the recurrent neural network which can effectively solve the problem of gradient vanishing or gradient explosion by introducing a set of memory units.

Anomaly detection in time series data is an important research direction. In view of the difficulties and challenges faced by anomaly detection, the method proposed in this paper is to let the LSTM neural network model learn the trend of the future time step, that is, to use the LSTM network as a predicted model. We use the stacked LSTM model to learn only normal time series data. Then the predicted error of the future time step is introduced into the Naive Bayes model [10] of Gaussian distribution to identify the abnormal behavior, as shown in Fig. 2.

Since LSTM has never been used in this set up, we first validate its utility and compare its performance to a set of strong baselines, that is, long short-term memory neural network (LSTM NN) and multi-layer perceptron model (MLP). The optimization goal of the stacked LSTM prediction model is to calculate only the losses in the final sequence step. At the same time, in order to improve the model generalization ability, we have used dropout [11] technology in the model training process, which further improves the performance of this model. The later experiments in this paper also prove that this is a more effective method.

The remainder of this paper is organized as follows: Section 2 deals with a review of time series processing. In Section 3, we describe the challenges and solutions for anomaly detection in time series processing. In Section 4, we use the stacked LSTM-Gauss-NBayes method and two contrastive methods LSTM NN model and MLP model for three real-world datasets, and the same time analyzing its results. Section 5 Summary and Future Outlook.

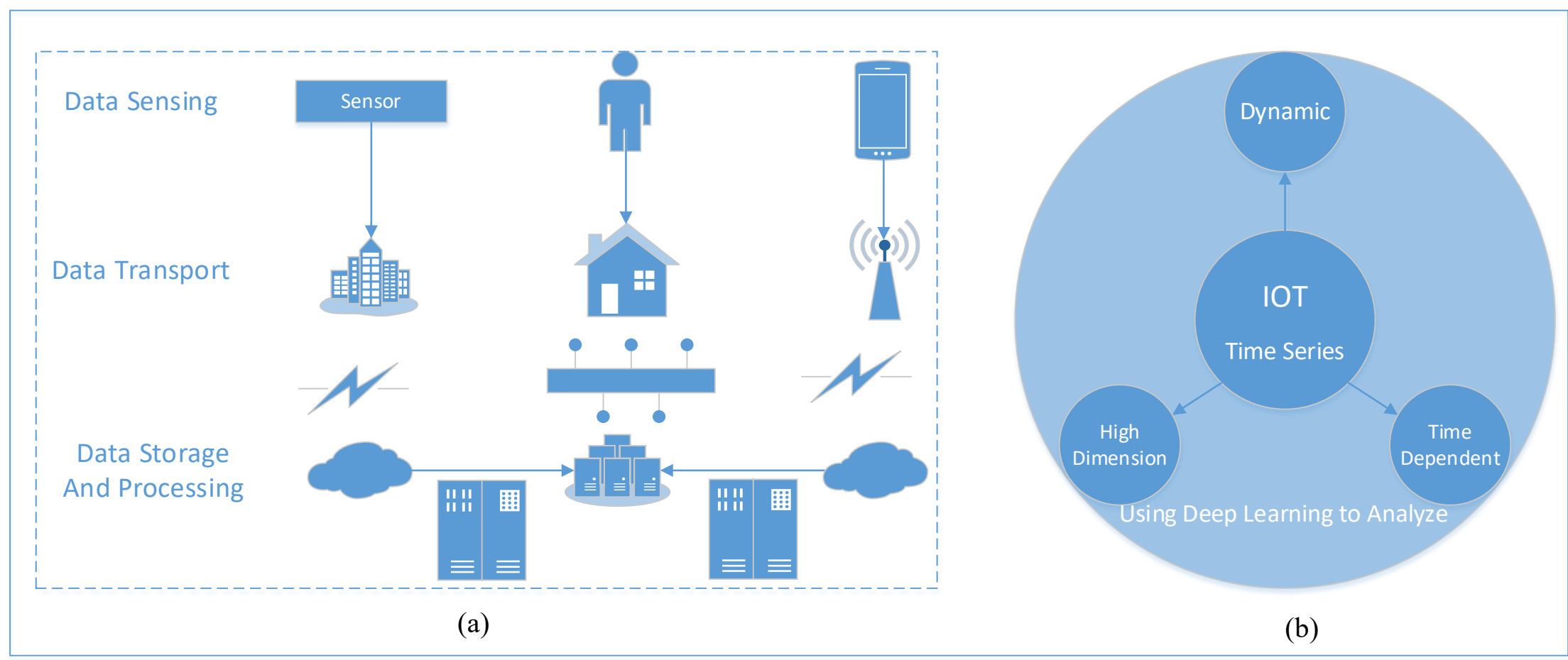

**Figure 1.** (a) shows the data collection process in a smart city, the sensor, human, mobile terminal generated data are sent to the cloud through such as switches, routers and other network equipment, and stored to the cloud server; (b) shows the characteristics of the time-series data of the Internet of things and use the deep learning to analyze it.

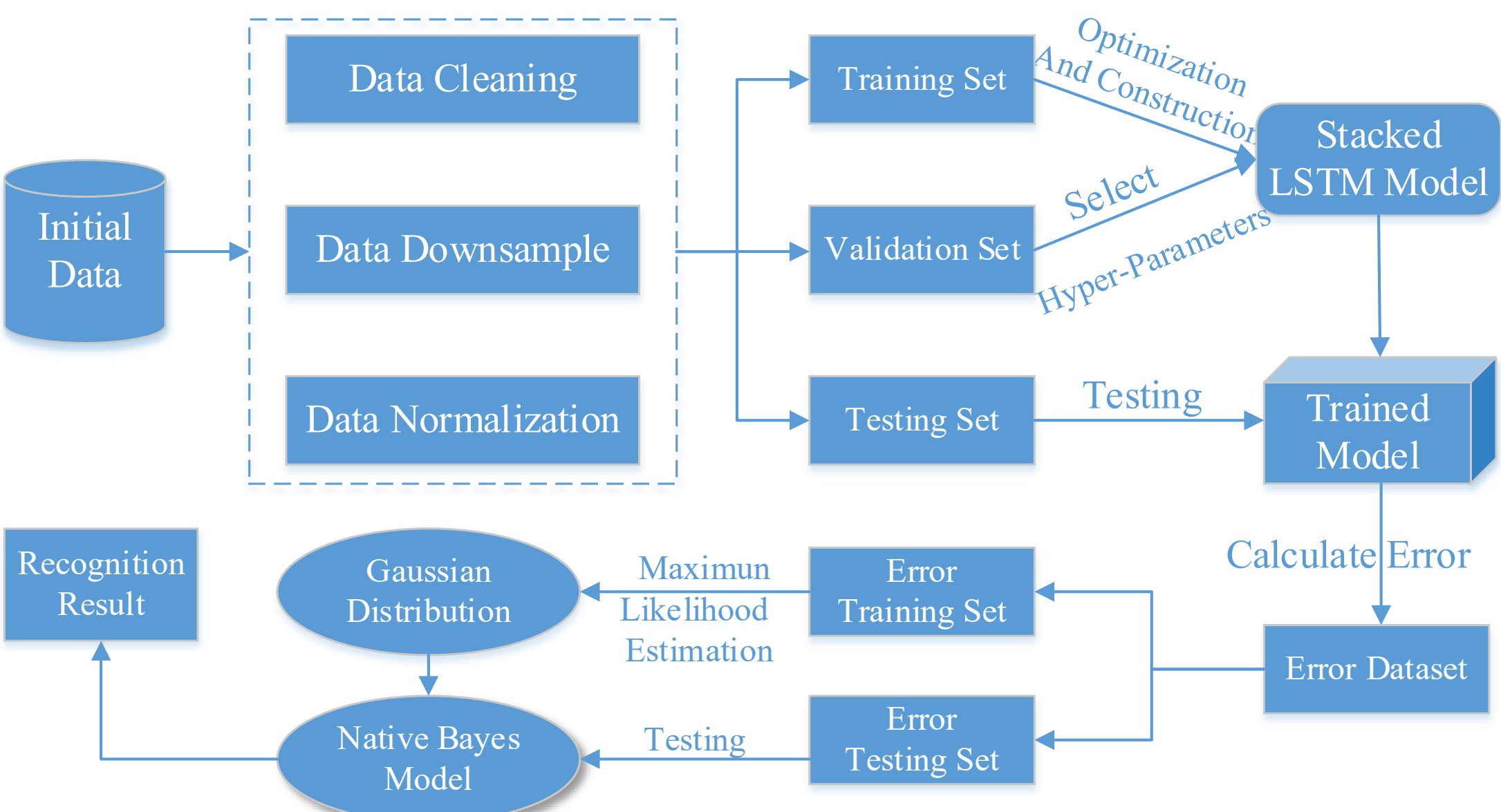

**Figure 2.** Construction procedure of anomaly detection model.

## 2 Development and Trend of Time Series Processing in IoT

### 2.1 The Application of Statistical Learning Method in Time Series Data

A time series is a sequence of numbers in chronological order. Traditional time series analysis uses mathematical statistical methods to analyze this sequence and predict the future development of things. Its basic principles are: first to detect continuous changes in the sequence. Application of past data to predict the development of trends. Next, the randomness of the time series is taken into account. Process effects may be affected by random factors, for which we use the weighted average method in statistical analysis to deal with historical data. A relatively representative model of the traditional time series method is the ARIMA model [12], which is called the Autoregressive Moving Average Model (ARIMA). ARIMA transforms the non-stationary time series into a stationary time series by d-order differential operation. Then the autocorrelation coefficient (ACF) and the partial autocorrelation coefficient (PACF) are obtained respectively for the stationary time series. Through the analysis of the autocorrelation graph and the partial autocorrelation graph, the optimal class $p$ and order $q$ are obtained. And then the ARIMA model is constructed from the above parameters. The method is simple and easy to master, but it has low accuracy and it is only suitable for short-term prediction. Time series prediction generally reflects three kinds of actual changes: trend change, cyclical change and randomness. The traditional time series analysis is commonly used in macroeconomic control of a national economy, regional comprehensive development planning, enterprise management and management, market potential forecast, meteorological forecast, hydrological forecasting, earthquake precursor forecast, crop pest disaster forecast, environmental pollution control, ecological balance, astronomy and oceanography.

### 2.2 Application of Neural Network in Internet of Things (IoT)

The neural network is a widely interconnected network of simple neurons which can adapt, and simulate the response of the biological nervous system to real world objects. A BP (Back Propagation) neural network has the ability to learn, memorize, associate, induce, generalize and extract features, tolerance faults and introspection. It can extract complex relationships between input and output, even when the relationship itself is in flux. Recently, BP neural network has been widely used to solve the problems of identification and prediction. It has achieved the effect that the conventional economics method cannot get in the economic field such as economic prosperity analysis, economic time series forecasting, portfolio securities optimization and stock forecasting. At the same time, BP neural network as a new time series prediction method can approximate nonlinear quantities with high prediction accuracy. Through the use of the time series relationships between before and later, the past observations as BP neural network's input and the future value as the BP network's output, which build a time series prediction model. From a mathematical point of view, The BP network becomes a nonlinear function of the input and output values.

**2.3 Application of Recurrent Neural Network in Internet of Things**

Nevertheless, although sometimes BP neural networks can achieve better results in time series processing, in traditional neural networks, we assume that all inputs (and outputs) are independent of each other. This is a very bad idea for many tasks. The Recurrent Neural Network (RNN) differs from the general feedforward BP neural network by memorizing the previous information and applying it to the calculation of the current output, that is, the nodes between the hidden layers are no longer connected. And the input of the hidden layer includes not only the output of the input layer but also the output of the hidden layer at the last time. Theoretically it is possible to process any length of sequence data. In practice, however, in order to reduce the complexity of the model, it is often assumed that the current state is only related to the previous several states. Although the simple recurrent neural network can theoretically establish a dependency between states of long time intervals, due to the simplicity of its hidden layer units' structure, the gradient explosion or gradient vanished is likely to occur in the relatively long-term time series processing task. This leads to the fact that only short-period dependencies can only be learned.

**2.4 Use the LSTM Neural Network to Process Time Series**

The long short-term memory neural network (LSTM) is a variant of the recurrent neural network, which can effectively solve the problem of gradient vanished or gradient explosion by introducing a set of memory units. It allows the network to learn when to forget the historical information of memory unit, when to update the memory unit with new information. At time $t$, the memory unit $c_t$ records all historical information up to the current moment and is controlled by three "gates": the input gate $i_t$, the forget gate $f_t$, and the output gate $o_t$, the elements' values of the three gates are set to [0, 1]. These models are well suited for data sets that contain time dimensions (such as web or server activity logs, sensor data from hardware or medical devices, financial transactions, or call records). Only the current state and some of the previous states are needed to train the network. LTSM can track dependencies and relationships across many time-steps. Although the use of typical feedforward neural network that receives the event window may also be done, the following window size will change along with the time. The feedforward method would limit us to the dependencies captured by the window, so the solution is not flexible. LSTM networks have been widely used in many sequence learning tasks. For example, give you a word sequence, we need to predict the likelihood of each word based on the previous word. LSTM language Models allow us to measure how likely a sentence is, which is an important part for Machine Translation (since high-probability sentences are typically correct). The LSTM network model can also predict the sequence of speech segments and their probabilities [13], given the input sequence of acoustic signals from sound waves. Other applications are handwriting recognition [14] and generating image descriptions [15].

## 3. Deep Learning for Anomaly Detection in IoT Data

### 3.1 Challenge in Time Series Anomaly Detection

Time series data has high dimensionality, complexity, dynamic, high noise characteristics, as shown in Fig. 1(b). If data mining is carried out directly on the original time series, it will not only spend a lot of resources and time in storage and computation, but also affect the accuracy and reliability of the algorithm. How to effectively preprocess the time series data under the condition that the key information of the time series data is not lost is a key problem. Reducing the dimensionality of the data and removing noise, are the key goals of preprocessing. Noisy data increases the complexity of an anomaly detection problem on the given time series data. At the same time, when the abnormal data is not available or sparse, it is difficult to learn the normal and abnormal sequence classification model.

Meanwhile, there is a certain recursive relationship between each point, and each event. There is no much value in the analysis of the single point. Therefore, time series data mining analysis needs to consider the logical relationship and the recursive relationship among the events. But when a time span is very large, which becomes a big challenge for the time series on the abnormal detection.

### 3.2 Use the Stacked LSTM-Gauss-NBayes Model to Detect Abnormal

For the challenges of time series data processing, we first use the down sampling technique to obtain the characteristic subsequence of the original time series. Down sampling reduces the number of dimensions in the original time series and makes it easier to learn patterns. At the same time, in order to speed up the convergence rate of the model, we normalize our data using min-max normalization for time series data, which is a linear transformation of the original data. The transformed values are mapped to the interval [0, 1].

Due to long-term and short-term dependency of time series in the Internet of things, we consider a LSTM neural network structure. In our LTSM, the input layer corresponds to a time series, the number of per hidden layer's LSTM units corresponds to the time step of the time series. And we use two hidden layers, that is, the stacked LSTM hidden layer (As shown in Fig. 3). For the output layer, we use the fully connected layer above the highest LSTM layer, followed by the element-shaped sigmoid activation function. We use the least squares loss function as the cost function for this model. Subsequently, we use the Back Propagation Through Time (BPTT) algorithm to train this model. For the data set, we divide the data set into a training set containing normal data, denoted as $D_{normal_train}$, a validation set containing normal data, denoted as $D_{normal_valid}$, a test set containing normal data, denoted as $D_{normal_test}$, and a test set containing abnormal data, denoted as $D_{abnormal_test}$. In the meantime, in the real Internet of things time series, abnormal samples are relatively small. We let the stacked LSTM prediction model only use the normal data set $D_{normal_train}$ to train, the hyper-parameters of which are determined by the validation set $D_{normal_valid}$. Furthermore, we put the test set $D_{normal_test}$ and $D_{abnormal_test}$ into the trained model respectively. The prediction results of the normal data and the abnormal data are obtained respectively. Then we can calculate the difference between the real data and the predicted data. So, we get the error data set, including the error of normal and abnormal data.

Next, we take the error at each point in each test sample ($D_{normal_test}$ and $D_{abnormal_test}$) as the attribute of the error data set. We divide the error data set into training sets $E_{train}$ and test sets $E_{test}$, where the labels values y belong to the set $\{0, 1\}$ and 1 for abnormal. So, we can build the Bernoulli distribution for the target value y and this distribution's parameters can be obtained by the training set $E_{train}$. For the meantime, we assume that each numerical attribute in the error data set is subject to the Gaussian distribution. In fact, this strong assumption is usually very effective and can produce robust results. After that, we establish the corresponding Gaussian probability density function for the conditional probability of each attribute. And the parameters of the Gaussian density probability function can be calculated by using the maximum likelihood estimation in the training set $E_{train}$. Then we use it to compute the conditional probability that the attribute occurs in the presence of a certain class. Due to the independence principle of Naive Bayes, we multiply these conditional probabilities of one sample directly. And we can get the conditional probability that one sample occurs in the presence of a certain class, that is to say, $P(x|y=0)$ and $P(x|y=1)$, where x, y stands for the sample and its label value respectively. Then, according to the Bayes formula, we calculate the posterior probability of the category of each sample in the test set $E_{test}$, i.e. whether this sample is abnormal. Because of the anomaly detection, we use the Precision and Recall and $F_\beta$ scores to measure our model. And β is greater than 0, which measures the relative importance of recall rate to precision rate. Here we use the $F_1$ score, Because of the data set referred to in this article, Precision and Recall is equally important.

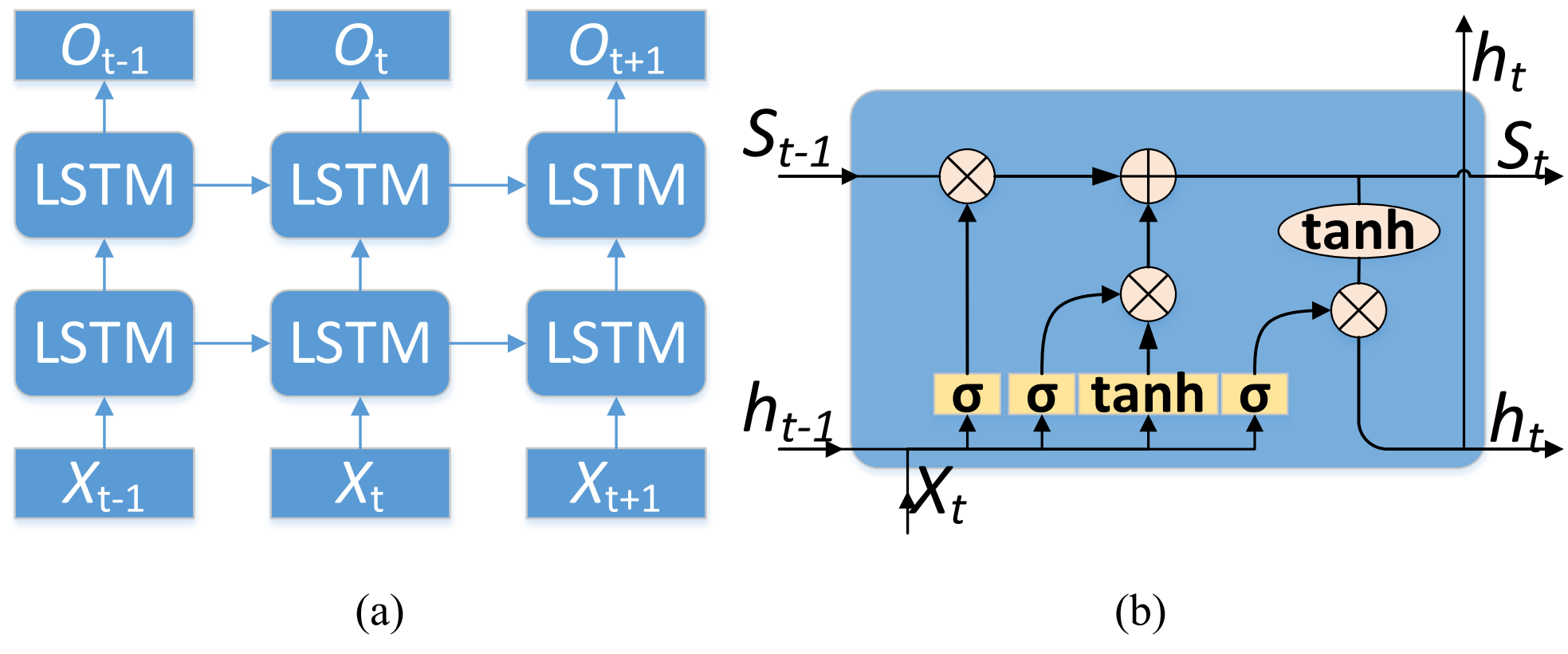

(a) (b)

**Figure 3.** Fig.3 (a) shows the expansion of the staked-LSTM prediction model, the LSTM units in the hidden layer are fully connected by recurrent connections. Each unit of the lower LSTM hidden layer in the stacked LSTM layer is connected to each cell in the LSTM hidden layer above it through the feedforward connection. In addition, Fig.3 (b) shows the internal structure of the LSTM layer, where σ and *tanh* represent the activation function. $X_t$ stand for the input of model. $h_t$, $h_{t-1}$ stand for the output of LSTM unit in the t-th sequence step and the previous sequence step respectively. $S_t$, $S_{t-1}$ stand for the value of LSTM memory unit in the t-th sequence step and the previous sequence step respectively.

## 4. Experiments

In this section, in order to demonstrate the effectiveness of our method. We use Google's deep learning platform, Tensorflow, to implement our algorithm and use the Nvidia GTX1070 to accelerate our training on the model. We did two sets of comparative method to validate our model, namely LSTM NN model and MLP model. For the LSTM NN model, we use an input layer, two hidden layers with LSTM memory blocks and a classification layer to construct it. Meanwhile, we use cross entropy loss function as the cost function for this model. For the MLP model, this is a conventional neural network model, constructed by an input layer, multiple hidden layers, and a classification layer. Here, we construct a multi-layer perceptron model with three hidden layers, each of which has 10, 20, 10 units respectively. And its cost function is also a cross entropy loss function. Furthermore, all models are trained on 80 percent of the data and tested on 10 percent. The remaining 10 percent is used as a validation set. We used the adaptive gradient algorithm (Adagrad) to train each model 1000 epoch. In order to prevent over-fitting, we use regularization techniques and dropout to reduce the complexity of these model. And we use 5-fold cross validation to select the hyper-parameters of these model. We consider three real-world data sets: power data, loop sensor data and land sensor data. The common characteristics of these data sets is that they are time series. Some of these data series are cyclical, and some are irregular.

**Table 1.** The autocorrelation coefficients of each data set at different delay cases. The autocorrelation coefficient is mainly used to describe quantitatively the relation between past events and current events.

| Dataset | Autocorrelation coefficient(ACF) | | | Time dependency |
|---|---|---|---|---|
| | Delay k=1 | Delay k=5 | Delay k=10 | |
| Power dataset | 0.79 | -0.78 | 0.56 | long |
| Loop sensor dataset | 0.71 | 0.40 | 0.05 | short |
| Land sensor dataset | 0.32 | 0.13 | 0.08 | very weak |

### 4.1 Data Set Description

**Power data**: This data is a user's power data for a year. It is collected every 15 minutes every day. We down-sample the original power data for each week, and the resulting data constitutes the input samples for our model. Under the normal circumstances, power consumption will be relatively high in the first 5 days of the week, and on the weekend, it will be relatively low. As we can see from Fig. 4 (a), the trend of power consumption shows 5 peaks in the first 5 days, and a trough appears two days later. If a sample had troughs in the first five days of a week, or wave crest appeared two days later, we could think of it as an anomaly. In addition, the data is noisy, so the peak does not appear exactly at the same time of the day.

**Loop sensor data set**: The data set is mainly composed of the number of vehicles passing through near the stadium and it was collected by the loop sensor. The data set was collected only when there was a game in the stadium. From the table 1, we can see that its ACF is less than 0.5 when the delay is 5, so we can know that this data set is a short-term time-dependent. In order to better analyze and use the data, we have screened the original time series, selected the data with only 1 hour before the game, the game and 2 hours after the game. From Fig. 4 (c), it can be found that the time series has a small peak in the first half and the second half, and the wavelet valley appears in the middle of the time series. Especially after the wavelet valley, vehicle data rises rapidly. This is quite consistent with a sharp increase in traffic after the race. Such behavior is considered normal for this time series.

**Land sensor data set**: The data set is mainly composed of land humidity data collected every 12 minutes by a sensor. Unlike the previous two data sets, they have some cyclical, but this data set is an irregular time series and very weak time-dependent data set, as shown in Table 1. With the development of time, this data set's values fluctuate randomly in a certain range. And the appearance of the anomaly is also irregular. So, it has increased the challenge to our model validation. In order to better analyze this data, similarly, we down-sampled the original time series, with 10 hours of humidity data as one sample, to train our model. It can be seen from Fig. 4 (f) that an anomaly may have occurred the first half and the second half of the time series.

**Table 2.** Experimental results, the performance of each model under four classification indicators (Accuracy, Precision, Recall, $F_1$) in three different data sets.

| Dataset | Method | Accuracy | Precision | Recall | F1 |
|---|---|---|---|---|---|
| Power dataset | LSTM-Gauss-NBayes | 0.969 | 1 | 0.941 | 0.962 |
| | LSTM NN | 0.905 | 0.846 | 0.931 | 0.886 |
| | MLP | 0.873 | 0.843 | 0.925 | 0.882 |
| Loop sensor dataset | LSTM-Gauss-NBayes | 0.953 | 0.932 | 0.976 | 0.954 |
| | LSTM NN | 0.870 | 0.867 | 0.897 | 0.881 |
| | MLP | 0.824 | 0.790 | 0.819 | 0.804 |
| Land sensor dataset | LSTM-Gauss-NBayes | 0.971 | 0.917 | 0.946 | 0.931 |
| | LSTM NN | 0.818 | 0.859 | 0.769 | 0.812 |
| | MLP | 0.818 | 0.889 | 0.727 | 0.800 |

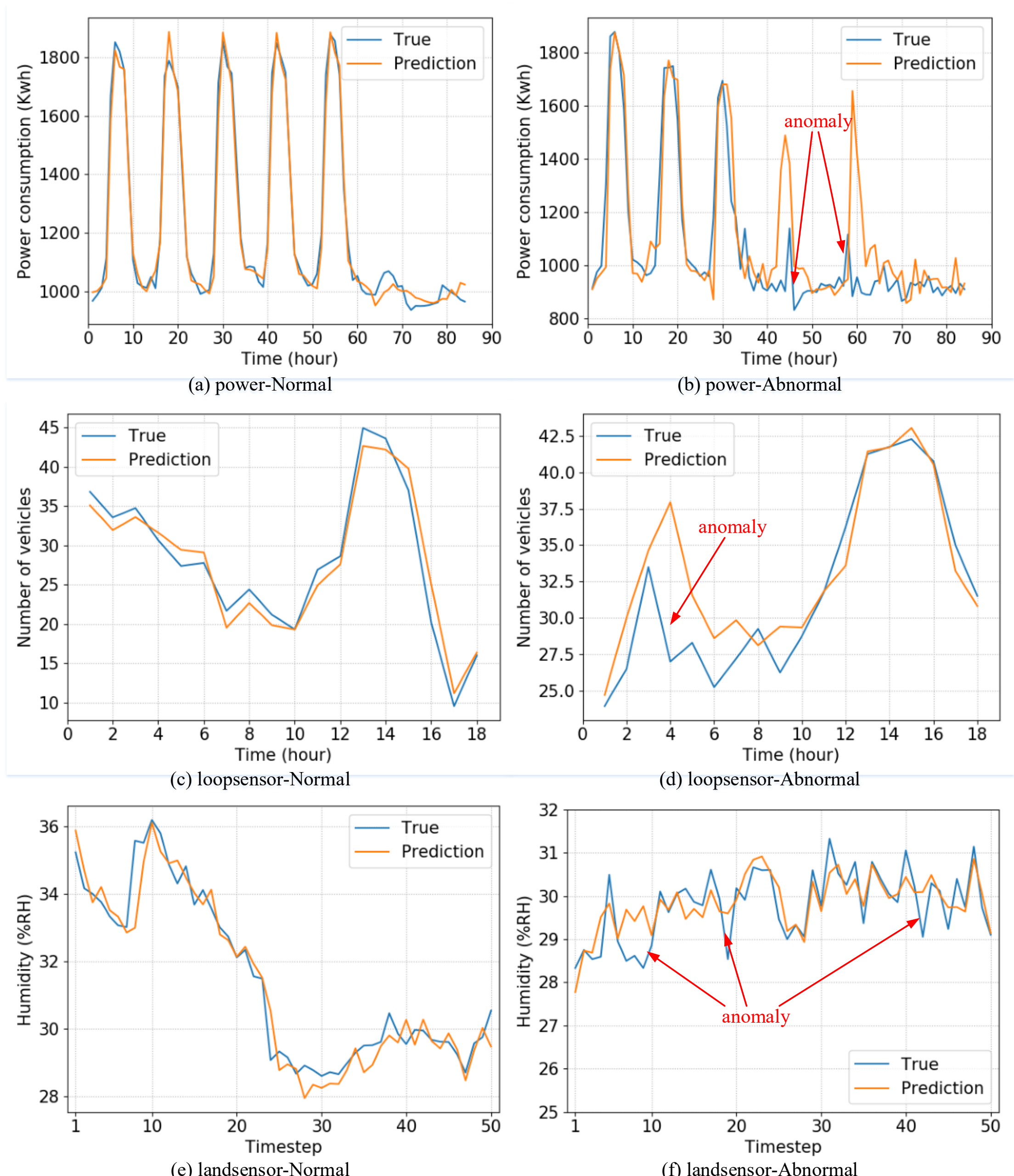

(a) power-Normal (b) power-Abnormal

(c) loopsensor-Normal (d) loopsensor-Abnormal

(e) landsensor-Normal (f) landsensor-Abnormal

**Figure 4.** The figure above shows the predicted results of the model under the three data sets, the orange lines represent the model predictions, and the blue lines represent the real result. In addition, the left side of the figure is the result of the normal sample, the right is the abnormal sample. The red arrow shows the possible anomaly.

## 4.2 Analysis and Comparison of Experimental Results

In order to evaluate our model, we use four indicators: accuracy, precision, recall, $F_1$ to comprehensively consider our model. Generally speaking, accuracy is enough to judge a model whose goal is to classify. Since our goal is to identify whether the sample is abnormal, precision and recall are important metrics by which to evaluate our model. Precision rate is mainly used to judge whether the classifier can correctly identify the anomaly, in other words, it mainly focuses on the identified abnormal samples of which how many such samples are really abnormal. And the recall rate is mainly judge whether the classifier can make all the abnormal samples identified. $F_\beta$ index is a combination of the previous two indicators, if β less than 1, then it represents the recall rate is more important. On the contrary, the precision rate has a greater impact on the model quality assessment. This experiment uses $F_1$ indicators, because for this article involved in the data set, precision rate and recall rate is equally important for us.

Table 2 shows the performance of each model under four classification indices in three different data sets. Under the same data set, we use the underscore to mark the highest value for each metric. For the power data set, we can see that its ACF is more than 0.5 when the delay is 10 from the table 1, so we can know that this is a data set with a long-term dependency on the time, and its current data is inextricably linked to the data in front of it. It can be seen from Table 2, that is why, the model with the hidden layer of the LSTM unit is generally better than the general hidden layer in the results of each of indicators. However, since the data set has a certain periodicity, the specific performance is that the electricity value in the first 5 days will be higher, and then two days later will be relatively low. Therefore, the feature extraction is also easy to implement for the ordinary neural network model. Thus, we can observe that the performance of the multi-layer perceptron model (MLP) is also good, its recall rate reached 88.2 percent.

For loop sensor data set, it is also closely related to the time order. However, compared to the power data set, the characteristics of this data set is not obvious, it is difficult to dig its feature, and it's difficult to distinguish between its abnormal time series and its normal time series. Therefore, it can be found from table 2, the result of two contrast experiments is unsatisfactory in this data set. Especially, the precision rate of anomaly detection for the multi-layer perceptron model (MLP), can be said to be bad, only reached 79 percent. Nevertheless, in this data set the LSTM-Gauss-NBayes model proposed in this paper is superior to the other two methods, and each of its indicators values are far ahead of the other methods' indicators values.

For the land sensor data set, this data set is not always dependent on the chronological order, in other words, it is difficult to predict future values given only past values. However, as can be seen from Table 2, our expected results are unexpectedly good, every indicator has reached a good result. From $F_1$ scores of Table 2, it can be seen that the long short-term memory neural network (LSTM NN) model and the multi-layer perceptron (MLP) model are less effective for the anomaly detection of this data set. Possibly this is because the features of the data set are not obvious and its value is within a frequently fluctuating range. In addition, it can also be found in Table 2, for this data set multi-layer perceptron (MLP) algorithm is better at precision rate of anomaly detection than the long short-term memory neural network (LSTM NN). This may be due to the fact that this time series does not exhibit strong time dependence.

## 5. Conclusion

In this paper, we propose a combination of long short-term memory neural network and Naive Bayes model with Gaussian distribution for anomaly detection. The LSTM neural network build model on normal time series behavior and then using the predictive error for the Naive Bayesian model of Gaussian distribution to detect anomalies. The LSTM-Gauss-NBayes method produces relatively good results on three real-world datasets, each of which involve long-term time-dependent and short-term time-dependent, even very weak time dependence. When compared with LSTM NN model and MLP model, our model always gives better or similar results, indicating that our model is robust. Future work may consider the use of multidimensional Gaussian distribution discriminant analysis instead of the Gaussian distribution of the naive Bayes model. This approach can take advantage of the relationship between the error data set properties since the covariance matrix is increased. In general, the LSTM-Gauss-NBayes method is a viable candidate in the field of anomaly detection.